\documentclass[preprint,showpacs,preprintnumbers,amsmath,amssymb]{revtex4}

\usepackage{graphicx}
\usepackage{dcolumn}
\usepackage{bm}

\begin{document}

\preprint{APS/123-QED}

\title{A Precursor of Market Crashes: \\ Empirical laws of Japan's internet bubble}

\author{Taisei Kaizoji}
 \altaffiliation{Division of Social Sciences, International Christian University, 3-10-2 Osawa, Mitaka, Tokyo 181-8585, Japan.}
\email{kaizoji@icu.ac.jp}

\date{\today}

\begin{abstract}
In this paper, we quantitatively investigate the properties of a statistical ensemble of stock prices. We focus attention on the relative price defined as $ X(t) = S(t)/S(0) $, where $ S(0) $ is the initial price. We selected approximately 3200 stocks traded on the Japanese Stock Exchange and formed a statistical ensemble of daily relative prices for each trading day in the 3-year period from January 4, 1999 to December 28, 2001, corresponding to the period in which the {\it internet Bubble} formed and {\it crashes} in the Japanese stock market. 
We found that the upper tail of the complementary cumulative distribution function of the ensemble of the relative prices in the high value of the price is well described by a power-law distribution, $ P(S>x) \sim x^{-\alpha} $, with an exponent that moves over time. Furthermore, we found that as the power-law exponents $ \alpha $ approached {\it two}, the bubble burst. It is reasonable to assume that when the power-law exponents approached {\it two}, it indicates the bubble is about to burst.  
\end{abstract}

\maketitle

\section{Introduction}
In the last several decades, large market crashes have frequently occurred in stock markets. The increasingly frequent market crashes have attracted the attention of academics, historians and policy makers. Numerous reports, commentaries, academic articles and books [1-10] have described the various reasons {\it why} market crashes have occurred. In the econophysisists' community, market crashes have also been a hot issue [11-20].  One of the greatest myths is that market crashes are random, unpredictable events. Can large market crashes be forecast? Do common warning signs exist, which characterize the end of a bubble and the start of a market crash? \par
Our previous work [19] quantitatively investigated the properties of an ensemble of {\it stock prices} in Japan in a 3-year period to study the warning phenomenon of large market crashes. We selected approximately $1200$ large stocks listed and traded on the Tokyo Stock Exchange for over than 20 years, and formed ensembles of daily stock prices in the 3-year period from January 4, 1999 to December 28, 2001, corresponding to the period of Japan's internet bubble and the market crash. We found that the tail of the complementary cumulative distribution function of the ensemble of stock prices in the high value range is well described by a power-law distribution, $ P(S>x) \sim x^{-\alpha} $, with an exponent that moves in the range of $ 1.09 < \alpha < 1.27 $. Furthermore, we found that as the power-law exponents $ \alpha $ approached {\it unity}, the internet bubble burst. This suggests that {\it Zipf's law} for an ensemble of stock prices is a sign of bursts of bubbles. In [20] we also quantitatively investigated the properties of an ensemble of {\it land prices} in Japan in the 22-year period from 1981 to 2002, corresponding to the period of the bubble forming and the crash in Japan's land market. We found the same laws that applied to the crashing of the internet bubbles held for the crash of Japan's internet bubbles. Namely, the tail of the complementary cumulative distribution function of the ensemble of land prices in the high price range is well described by a power-law distribution, $ P(S>x) \sim x^{-\alpha} $, and as the power-law exponents $ \alpha $ approached unity, the bubble collapsed in Japan's land market.\par 
As a result, our findings suggest that (i) the bubble phenomenon in asset markets is defined as the abnormal enlargement of price inequality which is caused by an over-concentration of investment capital, and (ii) that Zipf's law for ensembles of asset prices is an indication that a bubble is going to burst.\par
The aim of this paper is to attempt to extend our observation of Japan's internet bubble. In this paper, we use the daily-price data of almost all stocks (about 3200 stocks) listed on the stock exchanges in Japan in the 3-year period from January 4, 1999 to December 28, 2001, corresponding to the period of Japan's internet bubble and the crashes. We focus attention on investigating the statistical properties of ensembles of the {\it relative prices}, defined as $ X(t) = S(t)/S(0) $, where $ S(0) $ is the initial price, rather than those of the ensembles of the prices. 

\section{Internet bubbles}
In the 1990s, the personal computer, software, telecommunications and the internet were rapidly gaining acceptance for business and personal use. Computer-related technology drove the powerful bull market trend of global markets. Several economists even postulated that we were in a {\it New Economy}, where inflation was nonexistent and stock market crashes obsolete. Investors all over the world were euphoric, and believed in the fallacy of a perpetual bull market. Large scale stock speculation occurred, causing a worldwide mania. By the late 1990s, many technology companies were selling stock in initial public offerings (IPOs). Most initial shareholders, including employees, became millionaires overnight. By early 2000, reality began to sink in. Investors soon realized that the dot-com dream was really a bubble. Within months, the Nasdaq crashed from $5,000$ to $2,000$. In Japan's stock market, many {\it high-tech} stocks which were worth billions were off the map as quickly as they appeared. Panic selling ensued as investors lost trillions of dollars. {\it Yahoo Japan} is a typical example. Yahoo Japan's per share price was about 25,000 US dollars on January 5, 1998. After two years, the stock price surged by 5,450 percent, and reached a high of almost 1.5 million US dollars on February 23, 2000, and from then to December 30, 2002, the price declined to 15,000 US dollars, a drop of about 99.1 percent (See Figure 1). Numerous accounting scandals came to light, showing how many companies artificially inflated earnings, and many shareholders were crippled. Once again, we saw the development of a bubble and the inevitable stock market crash that always follows it [6-10]. \par
Figures 2(a) and 2(b) show respectively the movement of the mean and variance of the stock price ensemble distribution for each day in the three-year period from January 4, 1999 to December 28, 2001. The bursting internet bubble is explicit in the movement of the mean and variance. Both the mean and variance of stock prices abnormally surged until the beginning of 2000. Both the mean and variance peaked simultaneously on February 21, 2000, and began to drop rapidly. The reason why the abnormal enlargement of the variance of stock price occurred is probably because capital investment was extremely concentrated on stocks of industries which were related with personal computers, software, telecommunications and the internet, so that the prices of a tiny number of the large, high-tech stocks surged from the beginning of 1999. 
We can demonstrate this using the {\it Gini coefficient} $(G)$ [21], known as the index for wealth concentration. The Gini coefficient is defined as 
\begin{equation}
G=\frac{\sum^n_{i=1}\sum^n_{j=1}|S_i-S_j|}{2 \mu n^2}, 
\end{equation}
where $ n $ is the number of data and $ \mu $ is equal to the mean of the prices ensemble. Theoretically, the Gini coefficient ranges from zero, when all stocks are equal in price, to unity, when one stock has the highest price and the rest none. Figure 3 shows the movement of the Gini coefficient of the stock prices' ensemble. The Gini coefficient of the ensemble of stock prices approached {\it unity} on February 29, 2000, and price inequality reached the breaking point\footnote{If the distribution of the relative prices' ensemble follows the power law distribution, then the Gini coefficient can be written as $G=1/(2 \alpha - 1) $. Therefore, when the power-law exponent $ \alpha $ is equal to unity, the Gini coefficient is also equal to unity.}. 

\section{Empirical laws for ensembles of relative prices}
In this section, we will investigate the statistical properties of the ensemble of relative prices in order to study the statistical properties of cumulative price changes. The relative price is defined as 
\begin{equation}
X(t)=\frac{S(t)}{S(0)} 
\end{equation}
where $ S(0) $ is the stock price on the initial time $ t = 0 $, and is equal to the stock price on January 4, 1999. Taking the logarithm of $ X(t) $, we obtain the cumulative returns 
\begin{equation}
ln X(t)= \sum^t_{i=1} [ln S(i) - ln S(i-1)]. 
\end{equation}
Utilizing $ X(t) $ as an alternative to $S(t)$, we can investigate the statistical properties of cumulative price changes. For every relative price, $ X(0) $ for the initial time $ t = 0 $ is unity, so that the distribution of the relative prices for the initial time $ t = 0 $ is the delta function. We find that the tails of the complementary cumulative distribution functions of the relative prices ensemble in the high price range are well described by a power-law distribution, 
\begin{equation}
 P(X>x) \sim x^{-\alpha}.
\end{equation}
Figure 4(a) shows the log-log plot of the complementary cumulative distribution function $ P(X > x) $ of ensembles of the relative prices on February 23, 2000. The complementary cumulative distribution function is well approximated by a power-law distribution $ P(X > x) \sim x^{-\alpha} $ with $ \alpha = 1.93 $. The value of the exponent $\alpha$ is estimated by the ordinary least squares (OLS) regression in log-log coordinates. To confirm the robustness of the above analysis, we repeated this analysis for each of the trading days in the 3-year period from January 4, 1999, to December 28, 2001, which corresponds to the periods of Japan's internet bubble and its collapse. They are all consistent with a power-law asymptotic behavior. Figure 4(b) shows the movement of the power-law exponent $ \alpha $ in the period of Japan's internet bubble. The exponent $\alpha$ continued to decrease toward {\it two} during 1999, and the exponent $\alpha$ dipped below {\it two} from 17 February to 3 March, 2000, corresponding to the peak of the bubble and the onset time of the bursting of Japan's internet bubble. Then the exponent $\alpha$ started to go up. Figure 5(a) and 5(b) indicate the movement of the mean and variance of the prices ensemble distribution for each day during the 3-year period from January 4, 1999 to December 28, 2001. The bursting internet bubble is explicit in the movement of the variance of the relative prices rather than the movement of the mean of the relative prices. The variance of the relative prices abnormally surged until March 1, 2000 and then dropped sharply, while the dynamics of the mean of the relative prices do not indicate remarkable bursts of the bubble. These findings suggest that the threshold value of the exponent $\alpha$ that causes bubbles to burst is {\it two}. We interpret this as follows. The mean and variance of the power-law distribution (1) are written as 
$$ \mbox{Mean} = \frac{1}{\alpha - 1}; \quad 
\mbox{Variance} = \frac{\alpha^2}{(\alpha - 1)^2 (\alpha - 2)}. 
$$
Therefore, as the power-law exponent $ \alpha $ approaches {\it two}, the variance diverges so that the bubbles inevitably collapse. \par

\section{Concluding remarks}

In this paper, we focused our attention on the statistical properties of the ensembles of relative prices. Using about $3,200$ stocks traded on the Japanese Stock Exchange, we formed ensembles of daily relative prices $ X(t) $ in the period of Japan's internet bubble. We found that the tail of the complementary cumulative distribution function of the ensemble in the high range is well described by a power-law distribution, $ P(X>x) \sim x^{-\alpha} $, and when the power-law exponent $ \alpha $ approaches two, the bubbles of the relative prices collapse. Thus, it is possible to build up a hypothesis on bursting bubbles.\par
To date, no model so far has successfully explained the statistical properties of bursting bubbles. Hence, the next step is to model such behavior in stock markets.  A promising idea is to reproduce the statistical properties observed in the time evolution of financial prices by using the master equation or the Fokker-Plank equation\footnote{For an example of the modeling of market crashes, see [18].}. This study will be attempted in the future. 

\section{Acknowledgment} 
I wish to express my thanks to the members of financial engineering Group in Nikkei Media Marketing Inc, and Ms. Michiyo Kaizoji for their assistance in providing and analyzing the data. Financial support by the Japan Society for the Promotion of Science under the Grant-in-Aid, No. 15201038 is gratefully acknowledged. All remaining errors, of course, are mine.

\newpage

\begin{figure*}
\begin{center}
\includegraphics[height=14cm,width=12cm]{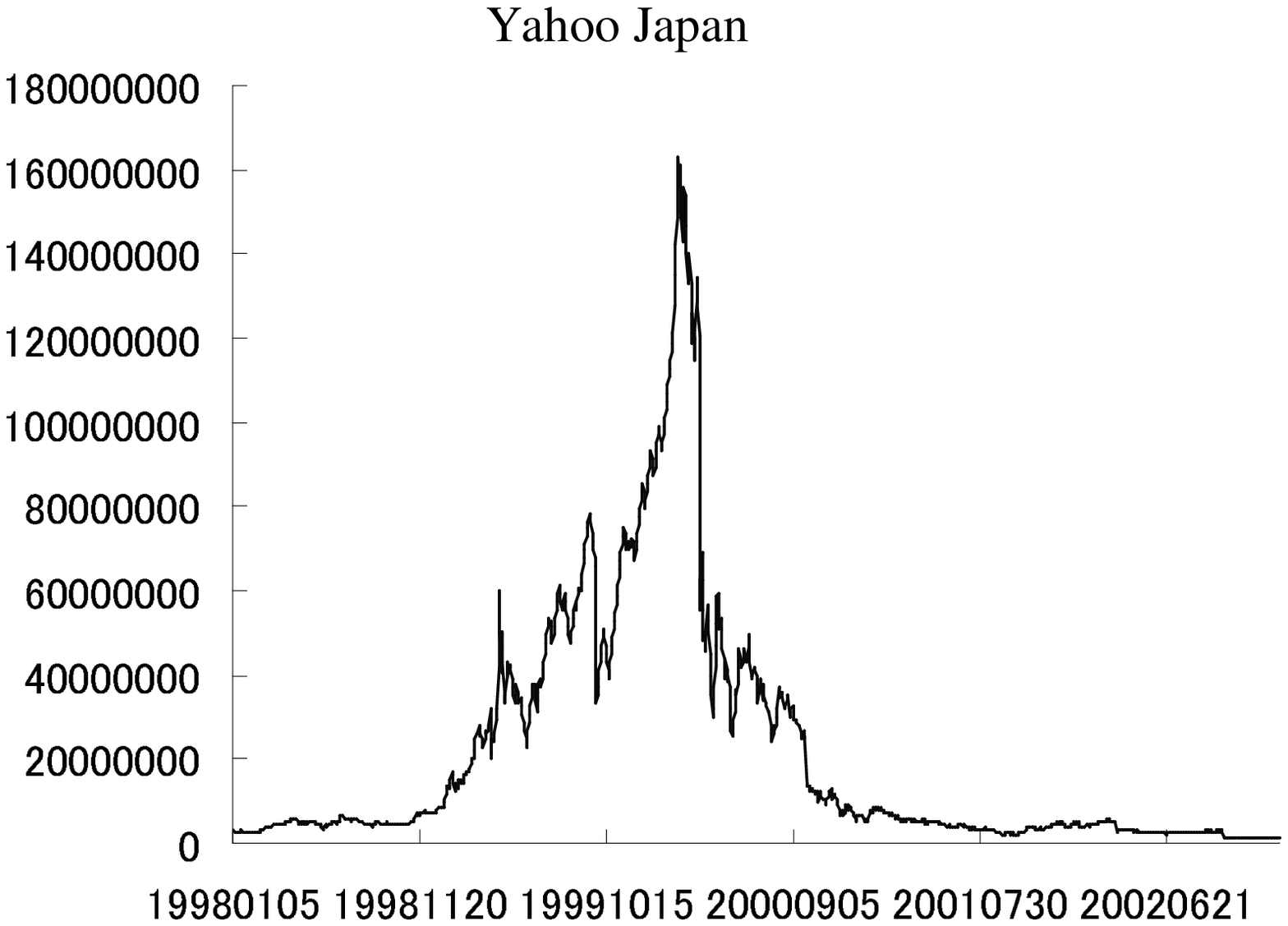}
\end{center}
\caption{The time series of the stock price of Yahoo Japan from January 5, 1998 to December 30, 2002. {\it Yahoo Japan} is a typical example of speculative bubbles ending in a crash.}
\label{fig1}
\end{figure*}

\begin{figure}
\begin{center}
  \includegraphics[height=14cm,width=12cm]{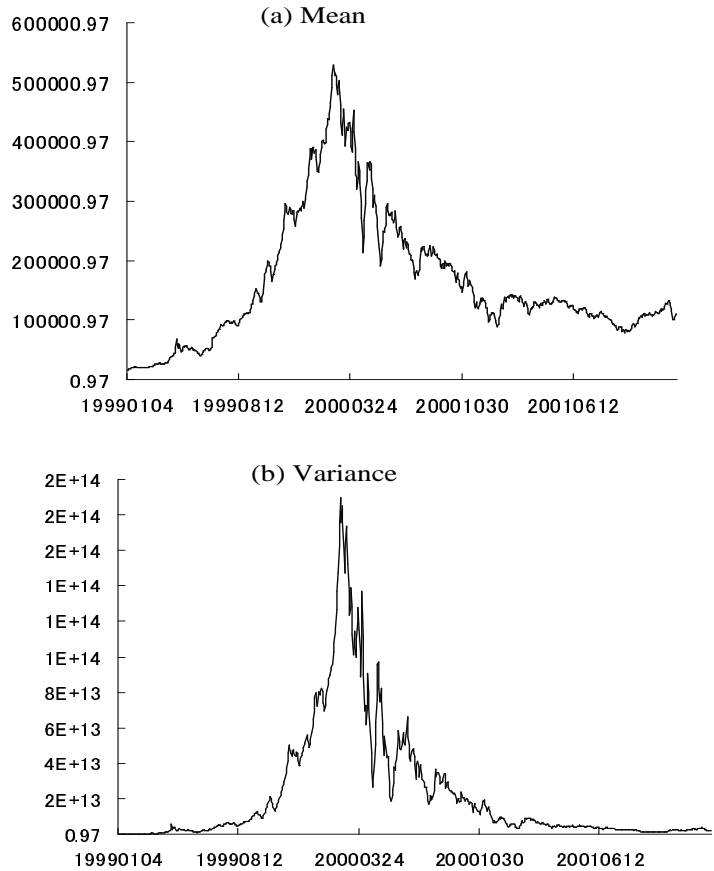}
\end{center}
\caption{(a) The movement of the mean of ensembles of the relative price $ S(t)/S(0) $ from January 4, 1999 to December 28, 2001, where $ S(0) $ is the stock price on January 4, 1999.  
(b) The movement of the variance of ensembles of the relative price from January 4, 1999 to December 28, 2001, where $ S(0) $ is the stock price on January 4, 1999. The stock prices' ensemble is composed by relative prices of about 3,200 Japanese companies listed on Japan's stock exchanges.}
\label{fig2}
\end{figure}

\begin{figure}
\begin{center}
  \includegraphics[height=14cm,width=12cm]{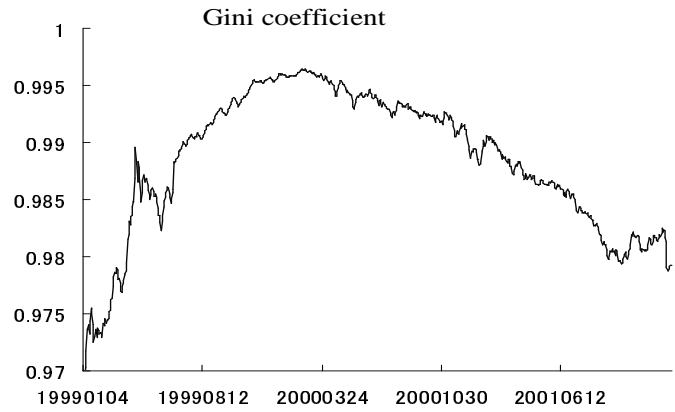}
\end{center}
\caption{The movement of the Gini Coefficient of ensembles of the stock price $ S(t) $ from January 4, 1999 to December 28, 2001. The stock prices' ensemble is composed of relative prices of 3,200 Japanese companies listed on Japan's stock exchanges.}
\label{fig3}
\end{figure}

\begin{figure}
\begin{center}
  \includegraphics[height=14cm,width=12cm]{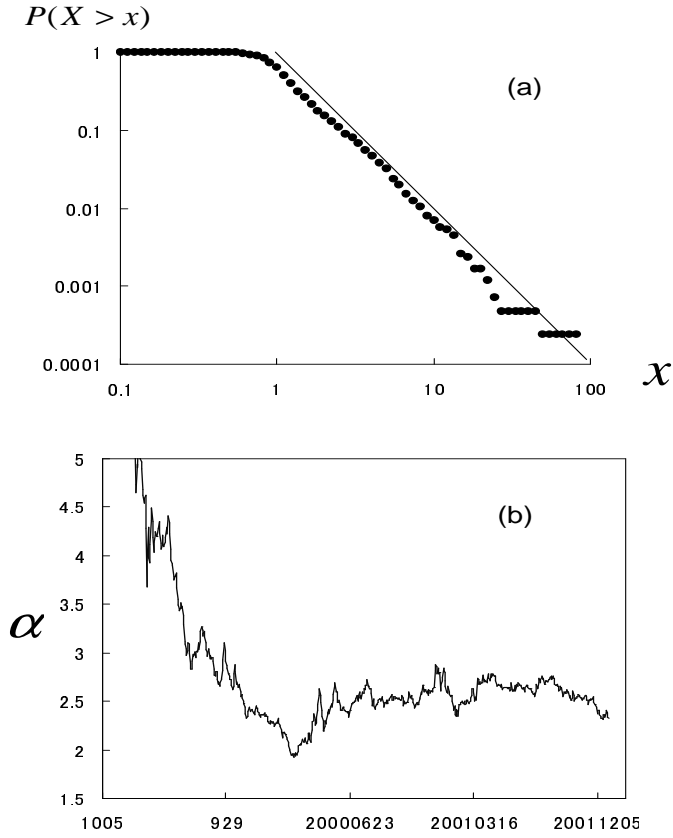}
\end{center}
\caption{(a) The log-log plot of the complementary cumulative distribution function $ P(X > x) $ of ensembles of the relative prices on February 23, 2000. The relative prices' ensemble is defined as $ X(t) = S(t)/S(0) $, where $ S(0) $ is the stock price on January 4, 1999. The complementary cumulative distribution function is well approximated by a power-law distribution $ P(X > x) \sim x^{-\alpha} $ with $ \alpha = 1.93 $. (b) The movement of the power-law exponent $ \alpha $ in the 3-year period from January 4, 1999 to December 28, 2001. The relative prices' ensemble is composed of relative prices of about 3,200 Japanese companies listed on Japan's stock exchanges.}
\label{fig4}
\end{figure}

\begin{figure}
\begin{center}
  \includegraphics[height=14cm,width=12cm]{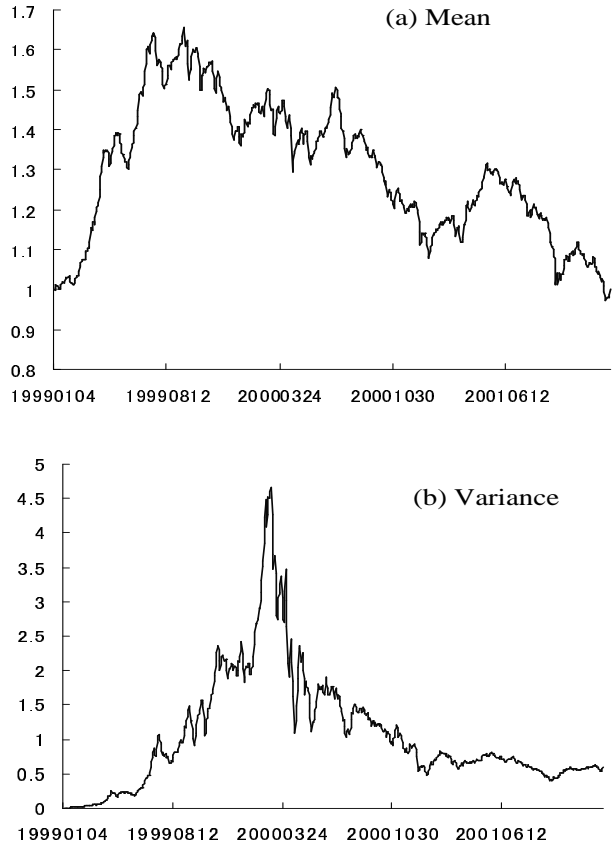}
\end{center}
\caption{(a) The movement of the mean of ensembles of the relative price $ X(t)=S(t)/S(0) $ from January 4, 1999 to December 28, 2001, where $ S(0) $ is the stock price on January 4, 1999.  
(b) The movement of the variance of ensembles of the relative price $ X(t) $ from January 4, 1999 to December 28, 2001, where $ S(0) $ is the stock price on January 4, 1999. The relative prices' ensemble is composed of the relative prices of about 3,200 Japanese companies listed on Japan's stock exchanges.}
\label{fig5}
\end{figure}
\end{document}